\documentclass[review]{elsarticle}
\usepackage{lineno,hyperref}
\usepackage{amsmath}
\usepackage{breqn}
\usepackage{calc}  
\usepackage{enumitem}  
\usepackage{amsfonts}
\usepackage{mathrsfs}
\usepackage{algpseudocode}
\usepackage{algorithm}
\usepackage{multirow}
\usepackage{bigstrut}
\usepackage{booktabs}
\usepackage{rotating}
\usepackage{hyperref}
\usepackage{bm}
\usepackage[top=3.5cm,bottom=3.5cm,left=3.5cm,right=3.5cm]{geometry}
\usepackage{caption}
\usepackage{subcaption}
\modulolinenumbers[5]
\usepackage{mathtools}
\usepackage{float}
\usepackage{adjustbox}
\usepackage{tikz}
\usepackage{dirtytalk}
\usepackage{tabularx}
\usepackage{lscape}
\usepackage{geometry}
\allowdisplaybreaks

\newcommand{\rev}[1]{{\color{black}#1}}

\usepackage{color}
\usepackage{lineno,hyperref}
\usepackage{amsmath}
\usepackage{breqn}
\usepackage{calc}  
\usepackage{enumitem}  
\usepackage{amsfonts}
\usepackage{mathrsfs}
\usepackage{algpseudocode}
\usepackage{multirow}
\usepackage{bigstrut}
\usepackage{booktabs}
\usepackage{rotating}
\usepackage{hyperref}
\usepackage[top=3.5cm,bottom=3.5cm,left=3.5cm,right=3.5cm]{geometry}
\usepackage{caption}
\usepackage{subcaption}
\usepackage{url}

\newcommand{\reals}{{I\kern-.35em R}}
\newcommand{\nats}{{I\kern-.35em N}}
\newcommand{\upto}{{\raise 1pt \hbox{$\scriptstyle \,\nearrow\,$}}}
\newcommand{\downto}{{\raise 1pt \hbox{$\scriptstyle \,\searrow\,$}}}
\newcommand{\eop}
	{\hfill{$\vcenter{\hrule height1pt \hbox{\vrule width1pt height5pt 
   	 \kern5pt \vrule width1pt} \hrule height1pt}$} \medskip}
\def\state #1. { \noindent{\bf#1.\enspace}}

 \usepackage{caption}

\journal{}

\biboptions{authoryear}

\makeatletter
\def\ps@pprintTitle{%
 \let\@oddhead\@empty
 \let\@evenhead\@empty
 \def\@oddfoot{}%
 \let\@evenfoot\@oddfoot}
\makeatother

\usepackage[]{acronym}
 
\newacro{gpio}  [GPIO]  {General-purpose input/output}
 
\begin{document}
\begin{frontmatter}

\title{A review on the charging station planning and fleet operation for electric freight vehicles}

\author[add1]{Md Rakibul Alam\corref{mycorrespondingauthor}}
\author[add1]{Zhaomiao Guo}

\cortext[mycorrespondingauthor]{(Corresponding Author) Ph.D. Candidate, 12800 Pegasus Drive, Orlando, FL 32816, Phone: 321-310-9950, Email: md894973@ucf.edu}

\address[add1]{Department of Civil, Environmental and Construction Engineering\\
Resilient, Intelligent, and Sustainable Energy Systems Cluster\\ 
University of Central Florida, FL 32766
}

\begin{abstract}
Freight electrification introduces new opportunities and challenges for planning and operation. Although research on charging infrastructure planning and operation is widely available for general electric vehicles, unique physical and operational characteristics of \acp{EFV} coupled with specific patterns of logistics require dedicated research. This paper presents a comprehensive literature review to gain a better understanding of the state-of-the-art research efforts related to planning (charging station siting and sizing) and operation (routing, charge scheduling, platoon scheduling, and fleet sizing) for EFVs. We classified the existing literature based on the research topics, innovations, methodologies, and solution approaches, and future research directions are identified. \rev{Different types of methodologies, such as heuristic, simulation, and mathematical programming approaches, were applied in the reviewed literature where mathematical models account for the majority. We further narrated the specific modeling considerations for different logistic patterns and research goals with proper reasoning. To solve the proposed models, different solution approaches, including exact algorithms, metaheuristic algorithms, and software simulation, were evaluated in terms of applicability, advantages, and disadvantages.} This paper helps to draw more attention to the planning and operation issues and solutions for freight electrification and facilitates future studies on \ac{EFV} to ensure a smooth transition to a clean freight system.

\end{abstract}


\end{frontmatter}

\newpage
\section{Introduction}\label{sec:intro}
\subsection{Background}

With the rapid development of e-commerce, freight \ac{VMT} has been rapidly increasing over the past decades \citep{jaller2022national} and is projected to further increase 57\% by 2049 \citep{vmt20222022}. Despite only accounting for 4\% of the total road vehicles in the U.S., freight vehicles have been the second largest contributors (23.6\%) of the \ac{GHG} emissions in the transportation sector, which has the highest share (28.7\%) of \ac{GHG} emissions in the U.S. \citep{usepa2019inventory}.

A significant percentage (45\%) of US ton-miles of freight is transported by trucks in 2020 and this percentage has kept increasing over the past decade. 
Due to the decarbonization trend in the power sector \footnote{21\% reduction since 1990, see \url{https://www.epa.gov/ghgemissions/sources-greenhouse-gas-emissions}}, \acp{EFV} could play a critical role to mitigate the \ac{GHG} emissions for  freight. 
The North American Council for Freight Efficiency suggests that the U.S. and Canada can convert more than 5 million medium- and heavy-duty trucks from fossil fuels to electricity without disrupting the freight flow \citep{promising2021}.

Recently, several auto manufacturers, such as Daimler, Mack, Tesla, Nikola, Volvo, Navistar, Rivian, and DAF, have announced plans to launch \acp{EFV} in the market \citep{volvo}. Most of the major delivery companies are also starting to expand their electric fleet. For example,  UPS has initiated a purchase for 10,000 electric delivery vehicles, while Amazon has committed to acquiring 100,000 \acp{EFV} through the emerging company Rivian. 
Additionally, in February 2021, the United States Postal Service granted a contract for the production and deployment of 50,000 to 165,000 EFVs over a span of 10 years according to a source\citep{usps}.

As the adoption of \acp{EV} in the freight industry is a promising trend, questions arise in terms of how to plan the supporting infrastructure and how to operate \acp{EFV} efficiently. First, due to lower energy efficiency and tighter time constraints, research on \ac{CS} planning is needed to mitigate the range anxiety and improve the operational feasibility of \acp{EFV}. \ac{CS} planning includes optimizing siting and sizing of \acp{CS}, the allocation of different types of charging plugs (i.e., \acp{EVSE}) at each \ac{CS} and feasibility study on the existing \ac{CS} network.
Second, research is also needed to better optimize the operational scheduling (i.e., the decision of charging, routing, platooning, and fleet sizing) of \acp{EFV} by incorporating the unique freight  characteristics (e.g., pickup and delivery time windows, driver's rest time, etc).  Although the previous research on \acp{EV} mainly focuses on light-duty \acp{EV}, such as private cars \citep{Qian2011,baghali2021impacts}, conventional taxis\citep{yang2017taxi}, and ride-sourcing vehicles \citep{alam2022charging,alam2022optimization}, some recent studies have looked into \acp{EFV} \citep{link2022technical,teoh2018decarbonisation,jahangir2021road,bac2021optimization}. A survey of the state-of-the-art research on  \ac{CS} planning and operational scheduling for \acp{EFV} will help to draw more attention to this field and facilitate future studies on \acp{EFV} to ensure a smooth transition to a clean freight system. 


%
\subsection{Goals and Contributions}

The goal of this review is to gain a better understanding of the state-of-the-art research on planning and operation of \acp{EFV}. In particular, we aim to answer the following research questions (RQs) in this review paper.

\begin{itemize}
    \item RQ1: What justifies the necessity for a dedicated research into \acp{EFV}?
    \item RQ2: What aspects of CS planning and operational scheduling were addressed in the existing literature??
    \item RQ3: What methodologies have been employed to tackle the planning and operational scheduling for \acp{EFV}?
    \item RQ4: How can we develop a fundamental mathematical programming model to address planning and operational scheduling for \acp{EFV}?
    \item RQ5: In what ways are various logistical configurations incorporated into modeling approaches for EFVs?
    \item RQ6: What solution approaches have been employed to address mall to large-scale problems, and how do they differ?
    \item RQ7: What are the potential research areas and challenges that have not been adequately explored?
\end{itemize}


Limited review papers are identified on the \ac{CS} planning, operational scheduling, and coupling of \ac{CS} planning and operational issues for \acp{EFV} considering different logistics stages (e.g., first-/last-/middle-miles). 
Up to November 2023, two review papers have been identified, which include selected topics on the operation of \acp{EFV}. For example, \cite{bektacs2019role} provided a review of the application of mathematical optimization on road and maritime freight systems, which included a brief discussion on the routing problem for \acp{EFV}. Similarly, \cite{patella2020adoption} reviewed literature on optimization, scheduling, policy, and sustainability issues for the last-mile logistics systems, where a section was devoted to routing and route-scheduling for \acp{EFV}. 

Our study offers a comprehensive examination of prior research related to \ac{CS} planning, operational scheduling, and the integration of \ac{CS} planning with operational considerations for \acp{EFV}. This review article can serve as a valuable resource for two distinct audiences. Firstly, researchers who are just embarking on their exploration of this field may benefit from this overview, which consolidates existing studies on EFV logistics across various freight patterns and may assist in identifying research gaps. For those interested in modeling specific aspects of the problem, this study can help identify previously employed modeling techniques and potentially highlight some modeling best practices. It also delves into the formulation of the most prevalent and favored approach, namely mathematical modeling, and provides insights into solving methods for such models. Secondly, decision-makers, such as EV infrastructure planners and logistics coordinators, may find this study to be a valuable resource for gaining insights into the current landscape of \ac{CS} planning and operational scheduling techniques. It has the potential to offer guidance on transition policies.



The remainder of this paper is organized as follows. Section \ref{sec:lit_search} provides a detailed description of our approach to retrieving relevant literature and presents analyses regarding the distribution of literature across journals, publication years, and the geographical locations of authors. Section \ref{sec:RQ1}-\ref{sec:RQ6} provide
answers to the aforementioned research questions by analyzing existing literature. Finally, Section \ref{sec:RQ7} provides conclusions and discusses future research directions (also the answer of last RQ).

\section{Literature Search}\label{sec:lit_search}
The following keywords and their combinations were used to guide the search process: ``freight electric vehicles”, ``electric truck”, ``green freight vehicles”, ``charging station allocation”, ``charging station planning", ``charging", ``routing”, ``scheduling”, and ``logistics". The following search engines were used: Science Direct, IEEE Explore, Web of Science, Scopus, Springer, and Google Scholar. Peer-reviewed journals, conference papers, and book chapters written in English were considered. After a review of the collected articles, we identified a total of 60 studies that were closely related to our study scope.


\begin{figure}[H]
	\centering
	\begin{subfigure}[t]{1\linewidth}
	\centering
        \includegraphics[width=0.7\linewidth]{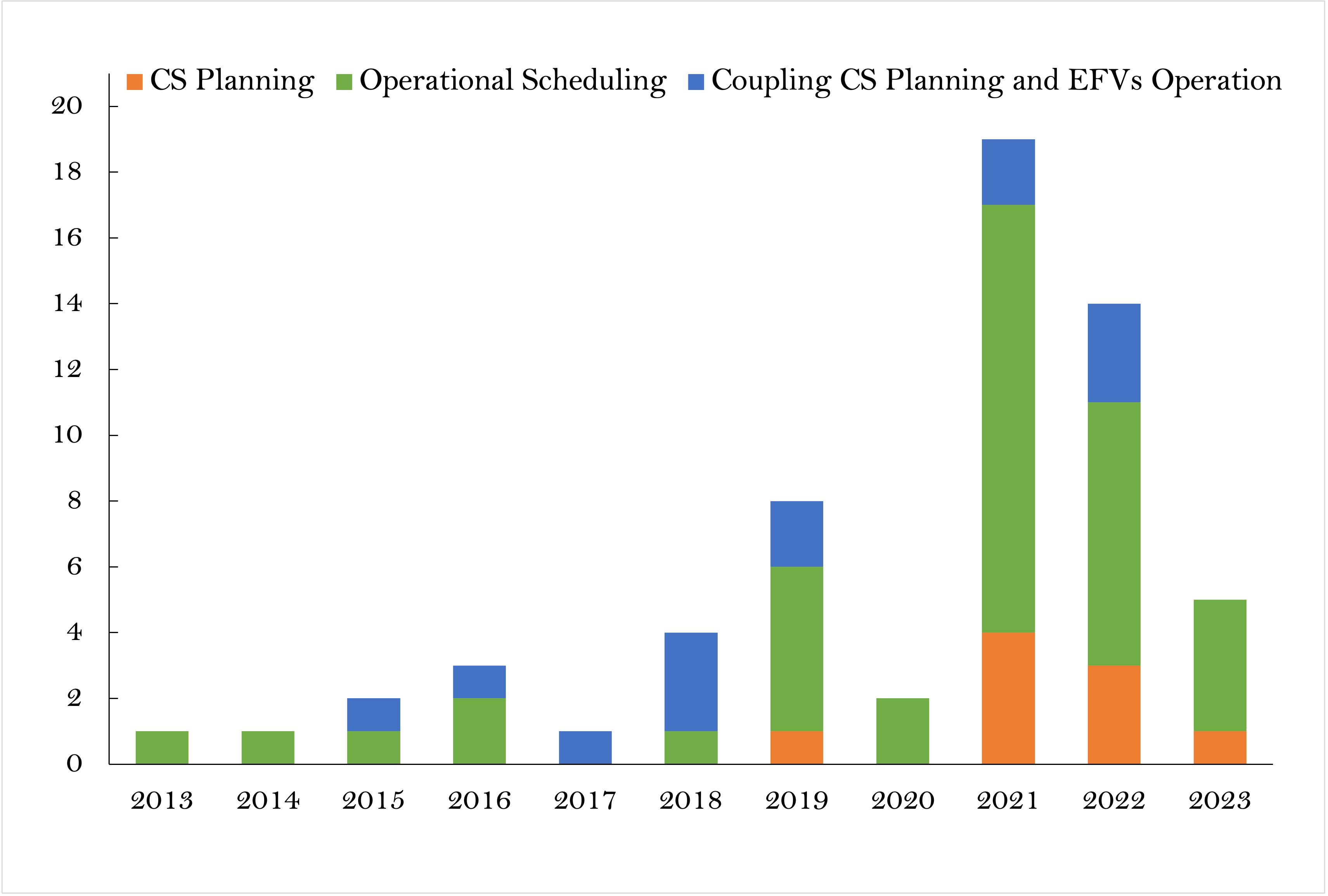}
	\caption{Trend of research publication per year}
	\label{fig:literature_year_content}
	\end{subfigure}
        \begin{subfigure}[t]{1\linewidth}
        \centering
	\includegraphics[width=0.7\linewidth]{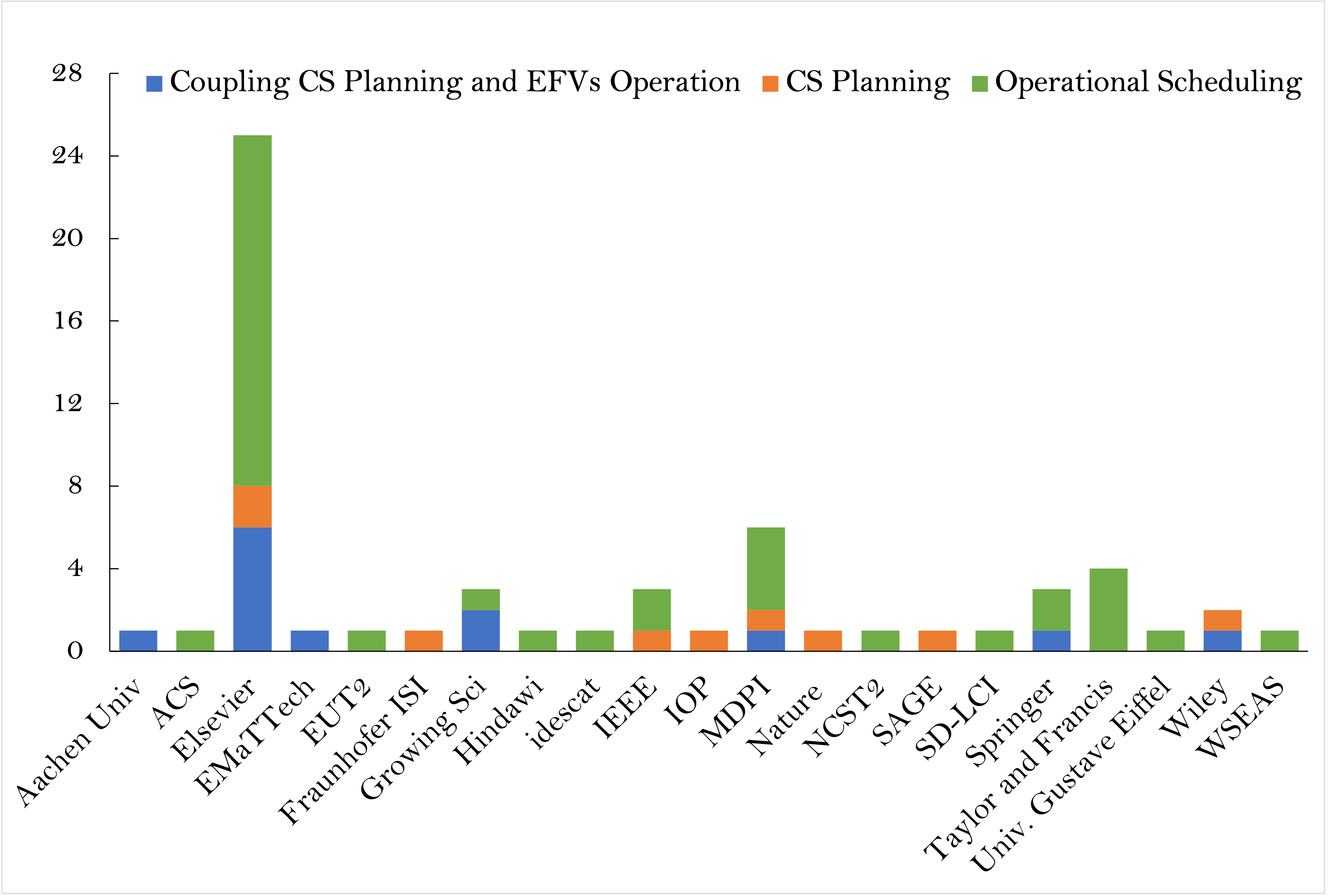}
        \caption{Research contribution from publishers}
        \label{fig:publisher_stat}
        \end{subfigure}
        \begin{subfigure}[t]{1\linewidth}
        \centering
	\includegraphics[width=0.7\linewidth]
        {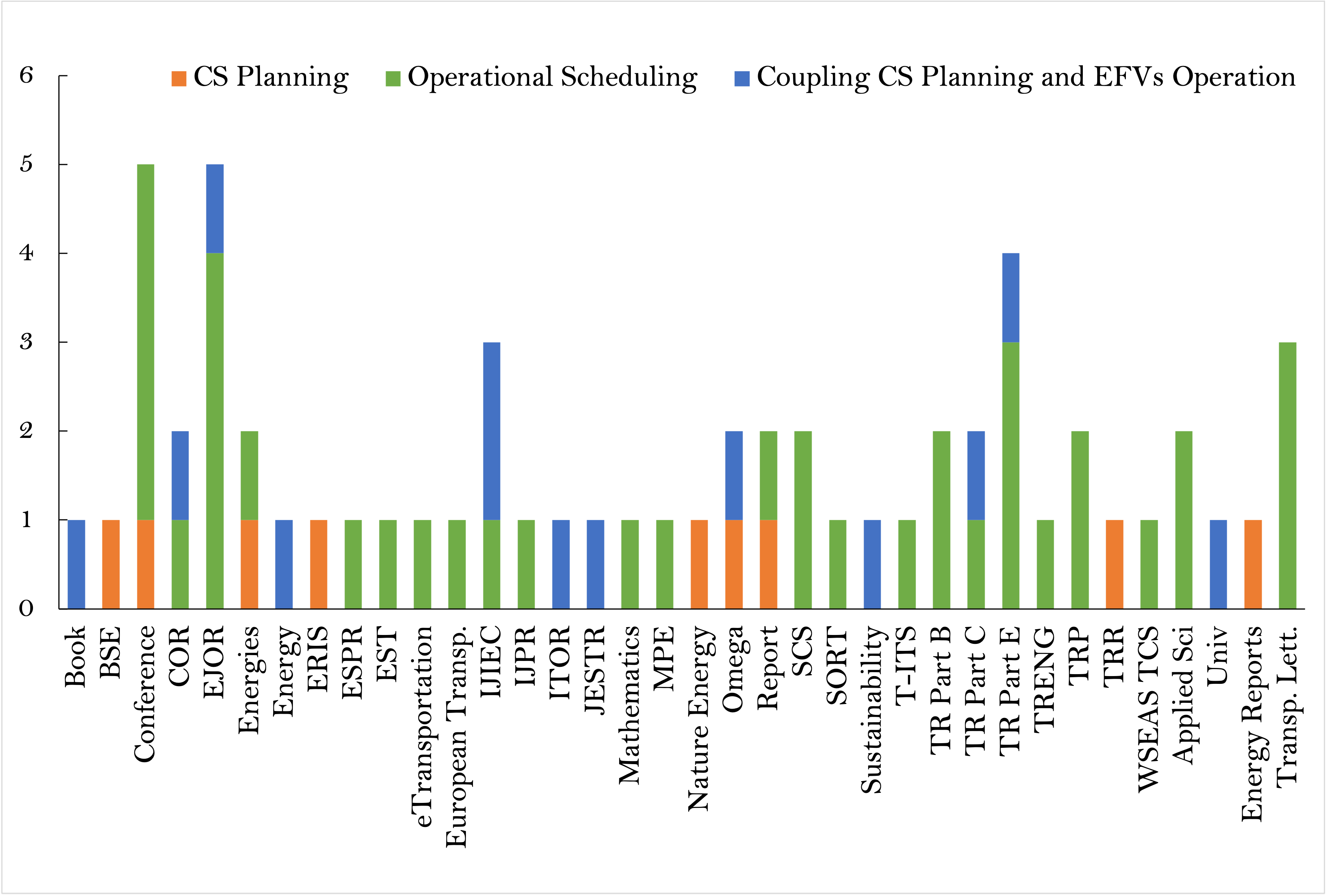}
        \caption{Research contribution from the journal, book, conference, etc.}
        \label{fig:journal_stat}
        \end{subfigure}
	\caption{Summary of literature search by year, publication type and publishers}
	\label{fig:summary_li_search}
\end{figure}

\begin{figure}[H]
	\centering
    \includegraphics[width=0.9\linewidth]{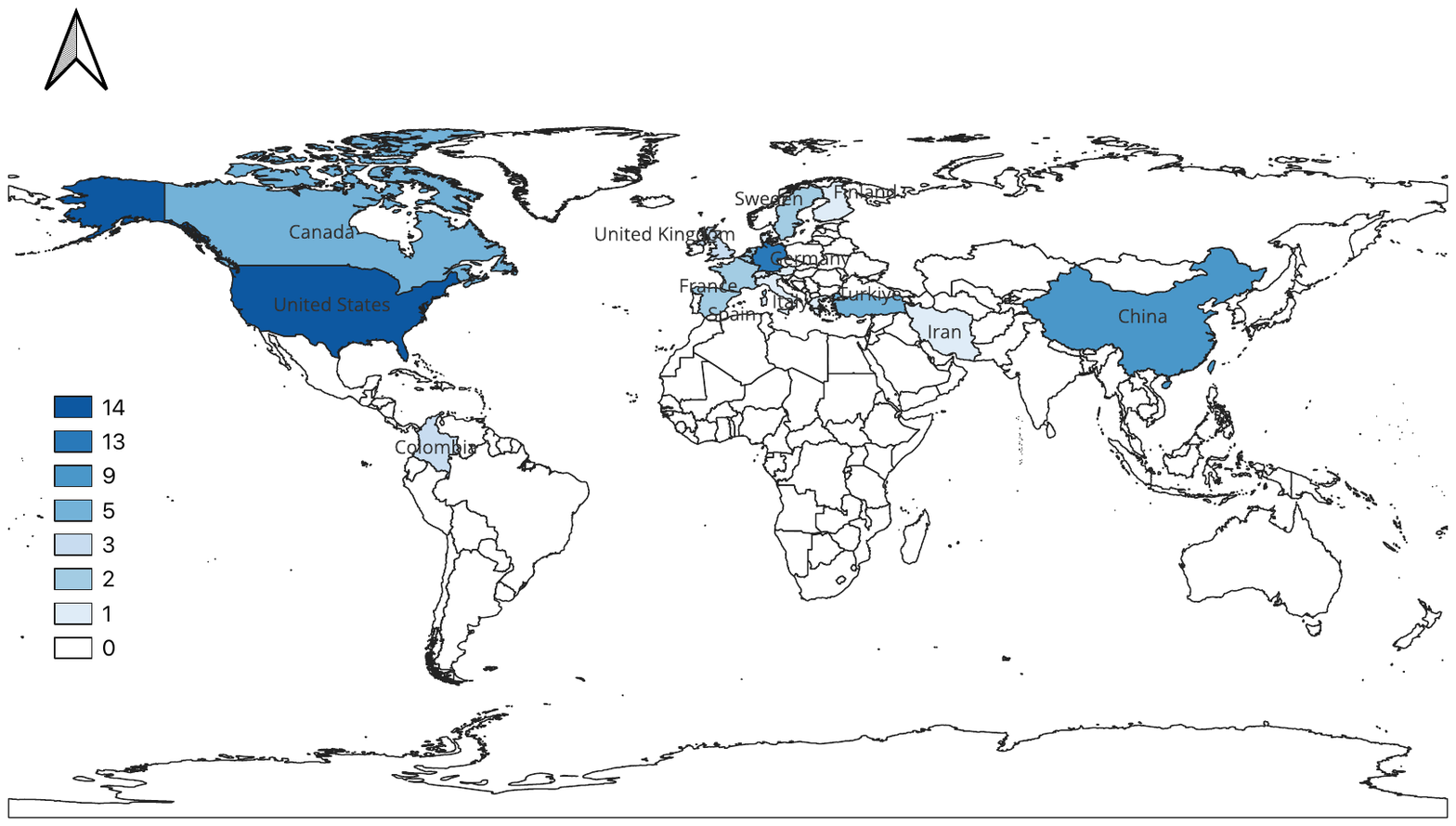}
	\caption{Geographic locations of existing literature}
	\label{fig:location_map_author}
\end{figure}

Figure \ref{fig:literature_year_content} presents the distribution of the collected literature by research focus and the year of publication. The total numbers of literature on \ac{CS} planning, operational scheduling, and coupling \ac{CS} planning and \ac{EFV} operation are 9, 38, and 13, respectively. This research field starts receiving more attention from the scientific community since 2018. Figure \ref{fig:publisher_stat} shows the distribution of the collected literature by publisher. The majority of the studies are published in Elsevier, with a total of 25 papers. MDPI published a total of 6, Taylor \& Francis published a total of 4, on the other hand, Springer, IEEE, and Growing Science published 3 articles each. The remaining publishers published one article each except Wiley (2 articles). Figure \ref{fig:journal_stat} presents the distribution of literature by journal names and publication types (i.e., conference proceedings, book chapters, and university repositories). The majority of these literature was published in the European Journal of Operational Research (with a total of five studies), Transportation Research Part E: Logistics and Transportation Review (with a total of four studies), Transportation Letters and International Journal of Industrial Engineering Computation (each with a total of three studies). Conferences published five pieces of literature. Literature with a focus on operational scheduling has a higher share compared with literature on \ac{CS} planning, as shown in Figure \ref{fig:summary_li_search}. Figure \ref{fig:location_map_author} shows that the United States and Germany emerge as prominent leaders in EFV research related to CS planning and operational scheduling, with China and Canada closely following suit.

\section{RQ1: What justifies the necessity for dedicated research into EFVs?}\label{sec:RQ1}

The travel patterns of freight vehicles differ significantly from those of passenger vehicles. Freight vehicle trips are more frequent and adhere to tight pickup and delivery schedules, in contrast to the less frequent and less rigorously scheduled trips made by passenger vehicles, with the exception of commercial passenger vehicles like ridesourcing and dial-a-ride services. The more frequent freight trips necessitate more charging due to lower energy efficiency resulting from the heavy freight payload.

Unlike private \ac{EV} users who have the flexibility to choose \ac{CS} with varying speeds, such as Level 1, Level 2, and fast chargers, based on their contingencies and specific time of day, \acp{EFV} require a dense distribution of \ac{DCFC} networks to meet tight freight schedules \cite{speth2022public}. While interstates and highways feature \ac{DCFC} stations, most of these are presently inadequate and cause challenges in accommodating and maneuvering large \ac{EFV} \cite{hurtado2021driving}. Furthermore, freight drivers require mandatory rest periods, particularly during long-distance middle-mile freight deliveries. These unique characteristics of \ac{EFV} demand special attention when planning CS infrastructure and operational scheduling for \ac{EFV}. In addition, different freight logistics patterns demand modifications in operational planning.

The extensive literature on \ac{EV} has primarily concentrated on passenger vehicles within the context of \ac{CS} planning\citep{alam2022optimization,cui2019electric,liu2019data,arslan2016benders,guo2016infrastructure,baouche2014efficient}. 
Unplanned CS allocation has been identified as a key hindrance to \ac{EV} adoption \cite{xu2022optimal} that can be equally true for \ac{EFV}. This issue can be exacerbated when operational scheduling for \acp{EFV} is implemented without considering the unique characteristics of these vehicles.


\section{RQ2: What aspects of CS planning and operational scheduling were addressed in the existing literature?}\label{sec:RQ2}
\subsection{Siting and sizing of \acp{CS}}\label{sec:siting_sizing}

The \ac{CS} siting and sizing problem decides the locations of \acp{CS} and the number of \acp{EVSE} at those \acp{CS}.
As electrifying freight vehicles is a new concept, few literature \citep{sauter2021charging,speth2022public} are found in recent days that considered unique operational characteristics of \ac{EFV} in \ac{CS} planning studies for \acp{EFV} . Flexible charging amount at \acp{DCFC} were considered and truck driver's break time (45-minute mandatory break after 4.5 hours of driving) was leveraged for en-route \ac{DCFC} charging. \cite{alp2022transitioning} partitioned the entire region into uniform-size grids with the center of each block to be potential \ac{DCFC} locations. The allocation of \acp{DCFC} would be done by freight industries annually to minimize both the investment costs (e.g., procurement cost and salvage value) and operational costs (e.g., electricity price, driver wage, maintenance of trucks and \acp{EVSE}). \rev{Nonetheless, the planning analyses presumed the immediate availability of land for the deployment of \acp{CS}, a condition that might not always be achievable, especially for large \acp{EFV} that require extra space for maneuvering. To reduce the infrastructure costs of \acp{CS},} \cite{bischoff2019impacts} considered the current gas stations as candidate \acp{DCFC} locations in Sweden and estimate the charging demand considering the variable charging speed as a function of \ac{SOC} and the queue at the \ac{CS}. 

\rev{Since the integration of electric transportation within the freight sector is in its early phases, the adverse effects stemming from the substantial charging load on the power grid have received limited attention in prior research on \acp{EFV}. However, with the increasing adoption of \acp{EFV}, the impacts of their charging load on the power grid can be significant due to much higher charging power. As a result, there exists a need for a well-grounded \ac{CS} planning strategy that takes into account the interconnected nature of transportation and power requirements to enhance efficiency and reliability for \ac{EFV} charging.} Recently \cite{li2023centralized} conducted a study that decides the optimal number of \acp{EVSE} of different charging power and the upgrade plan of power distribution systems considering two  charging strategies: unmanaged (immediate charging without central coordination of power supply) and managed charging (smart charging with central coordination of power supply).    

\subsection{Feasibility study for existing \ac{CS}}\label{sec:feasibility_study}

The feasibility study aims to evaluate if a certain charging infrastructure deployment plan is feasible to support \acp{EFV}' charging needs. \cite{hurtado2021driving} assessed the maximum area that can be served by \acp{EFV} considering \acp{DCFC} at truck stops (i.e., resting and parking facilities) along the US-Interstate Highway Network. It is found that service coverage is less for the states located west of the Mississippi River (excluding California, Oregon, and Washington) due to sparser road networks and fewer truck stops. \cite{borlaug2021heavy} assessed the potential impacts of depot charging on the electric grid during an average of 14 hours of off-shift dwell time for heavy-duty \acp{EFV} considering three charging strategies: charge at 100 kW immediately (EVs charged as soon as possible), charge at 100 kW delayed (EVs charged as late as possible), and charge at a constant minimum power (EVs charged as slowly as possible). The data was based on the National Renewable Energy Laboratory (NREL) Fleet DNA database. Their study found that the existing power distribution systems can meet the \ac{EFV} charging demand with 100kW \acp{EVSE}. \cite{jahangir2021road} conducted a study to measure the range anxiety of medium and heavy-duty \acp{EFV} under the existing \ac{CS} network in Finland and Switzerland. They found that the current \ac{CS} network (consists of minimum 40kW \acp{CS}) is capable to serve 93\% and 89\% of trip mileages in Finland and Switzerland, respectively, for \acp{EFV} with a maximum of 30-ton payload capacity. \rev{However, the \acp{CS} within the depots and vendor-owned \acp{CS} (e.g., Tesla Supercharger) only serve the proprietary \acp{EFV}, which can be resolved by proper charging sharing policy. Therefore, }an economic-feasibility study considering different business models of sharing existing \ac{CS} in Stockholm, Sweden was also conducted by \cite{melander2022benefits}. It was found that sharing \ac{CS} policy increases utilization of \acp{EVSE} and enables more environment-friendly freight transport.    

\subsection{Routing problem}
Routing problems are crucial in the freight industry with \acp{EFV} especially considering the limited driving range, sparse charging stations, and tight delivery schedules. This decides the transportation
links to be traversed by EFVs and the time of traversing these edges for freight delivery. The classic routing algorithms such as Dijkstra's algorithm or Bellman-ford algorithm cannot be directly applied because the ``edge cost'' is negative during the charging event at \ac{CS}, whereas the cost is positive while traveling (e.g., energy consumption and travel time) \citep{ioannou2020freight}.

Besides the unique physical and operational characteristics of \acp{EFV}, different patterns of logistics and the consideration of various important factors such as energy consumption uncertainty, stochastic travel time, stochastic battery depletion and a mixture of fleets generate different variants of \ac{VRP}.  

For example, \cite{afroditi2014electric,napoli2021freight,wang2021integrated} studied \ac{EVRPTW} considering homogeneous \ac{EFV} fleet, whereas \cite{celebi2021planning,goeke2015routing} studied \ac{MVRPTW} considering a mixture of \acp{EFV} and \acp{ICEV}.  

\cite{zhao2021vehicle} tried to optimize the \ac{EFV} dispatch and routing for dedicated-service (pickup and delivery) logistics considering delivery time windows, flexible charging, dynamic energy consumption due to freight load change, and dynamic traffic situation. The routing problem was named \ac{EVRP-PD-TW-PR}. 

Although most studies consider fixed delivery locations, \cite{sadati2022electric} introduced flexible freight delivery locations in \ac{EVRP-FD} given delivery time windows. All of these studies considered customer locations for either pickup or freight delivery, \cite{yilmaz2022variable} provided flexibility for both pickup and delivery at the same customer location by introducing \ac{SPD} in \ac{EVRP-SPD}. 

Some studies of routing problems also investigate the potential charging load in the power distribution network and the consumption of energy during traveling by energy model. \cite{ahmadi2021vehicle} investigated the importance of vehicle dynamic parameters (e.g., road gradient, acceleration/deceleration, and extra mass) in the energy model for \ac{EVRP-PD}. \cite{arias2021linearized} studied \ac[2]{eLCVRP-CS-PDS} for \acp{EFV} where the \acp{CS} were connected at power distribution nodes. The uncertainty of energy consumption was addressed by \cite{PELLETIER2019225} in \ac{EVRP-ECU}.

Uncertain behavior of traffic parameters such as traffic volume and travel time is also considered in a few studies of routing problems. For example, \cite{zhao2020distribution} emulated the real traffic situation in \ac{EVRP-TTC}. Similarly, the uncertainty of travel time was addressed by \cite{reyes2019simheuristic} in \ac{EVRP-ST}.

Unlike routing only for plug-in \acp{CS}, \cite{fan2022half} proposed a more generalized \ac{HOTDMD-EVRP-BRS} approach where \acp{EFV} have charging opportunities in both plug-in \acp{CS} and \acp{BSS}.

\subsection{Charge scheduling}\label{sec:charge_scheduling}
Only the decision of charging locations from \ac{CS} planning is not sufficient, efficient charge scheduling contributes to successful freight operation. Charge scheduling decides the charging location among \acp{CS}, time to charge, and charging amount. Based on charging location, charging scheduling studies can be categorized into three categories: charging at depot \citep{al2022smart,klein2022branch}, charging at both depots and en-route \acp{CS} \citep{woody2021charging,sadati2022electric,akbay2022variable,ahmadi2021vehicle,afroditi2014electric,bac2021optimization,goeke2015routing,goeke2019granular} and charging at customer's locations (e.g., \citep{elahi2022modeling}). 

Full charging at \ac{CS} \citep{andelmin2019multi,yilmaz2022variable} causes a huge peak load that necessitates the upgrade of the power supply, whereas charge scheduling considering the flexibility of charging quantity or mixed charging rate coordinated with the power supply can avoid the additional cost of power system upgrade. To justify this statement, \cite{woody2021charging} investigated four charging behaviors, including immediate-full charging, delayed-full charging, immediate-sufficient charging, and delayed-sufficient charging. These charging behaviors were characterized by departure/arrival time (e.g., charging \acp{EFV} to target SOC right before departure in delayed charging; charging \acp{EFV} immediately after arrival in immediate charging) and battery \ac{SOC} (e.g., full charging before departure from depot; sufficient charging to complete daily trips before departure). To resolve such issue, \cite{al2022smart} proposed a smart charging system at commercial facilities (depot only) for heavy-duty \acp{EFV} considering \ac{OSPT}, \ac{IBDR}, and \ac{V2G} technologies. Similarly, \cite{elahi2022modeling} considered charging scheduling for \acp{CS} with \acp{EVSE} with mixed charging rate.

\subsection{Platoon scheduling}
Although the long charging duration of \acp{EFV} may delay the shipment, and the platooning requires perfect coordination between trucks’ travel schedules over space and time, there is a great opportunity of saving energy consumption if these two technologies can be co-optimized in platoon scheduling that decides the time and location when EFVs form platoons along the route. On one hand, the energy saving by platooning will significantly reduce the need for en-route charging and avoid shipment delays. On the other hand, \acp{EFV}’ charging activities provide flexibility for them to coordinate their spatial-temporal trajectories, which offers more opportunities for \acp{EFV} to platoon \citep{ALAM2023104009}. 

\cite{scholl2023platooning} optimized the total cost of energy consumption considering platoon position using meta-heuristic approaches whereas \cite{ALAM2023104009} minimized the total cost of delivery delay, hub charging, and en route charging considering the charging station capacity and solved the problem by CPLEX. These studies consider different platooning settings and objectives. For example, \cite{scholl2023platooning} proposed platoon formation at service points (some service points have charging facilities) along the route where platoons are formed among waiting trucks. On the other hand, \cite{ALAM2023104009} proposed platoon formation at \acp{CS} with consideration of \ac{CS} capacity and driver's rest time. 



\subsection{Coupling \ac{CS} planning and \acp{EFV} operation}\label{sec:coupling_allocation_route}

While optimizing the \ac{CS} siting and sizing, some studies explicitly model \ac{EFV} routing to better reflect the operational performance. These decision problems are termed as \ac{LRP}, which integrates the \ac{EVRP} and the \ac{CS} allocation problem in a single modeling framework.  

\begin{figure}[H]
    \centering
    \includegraphics[width=0.75\linewidth]{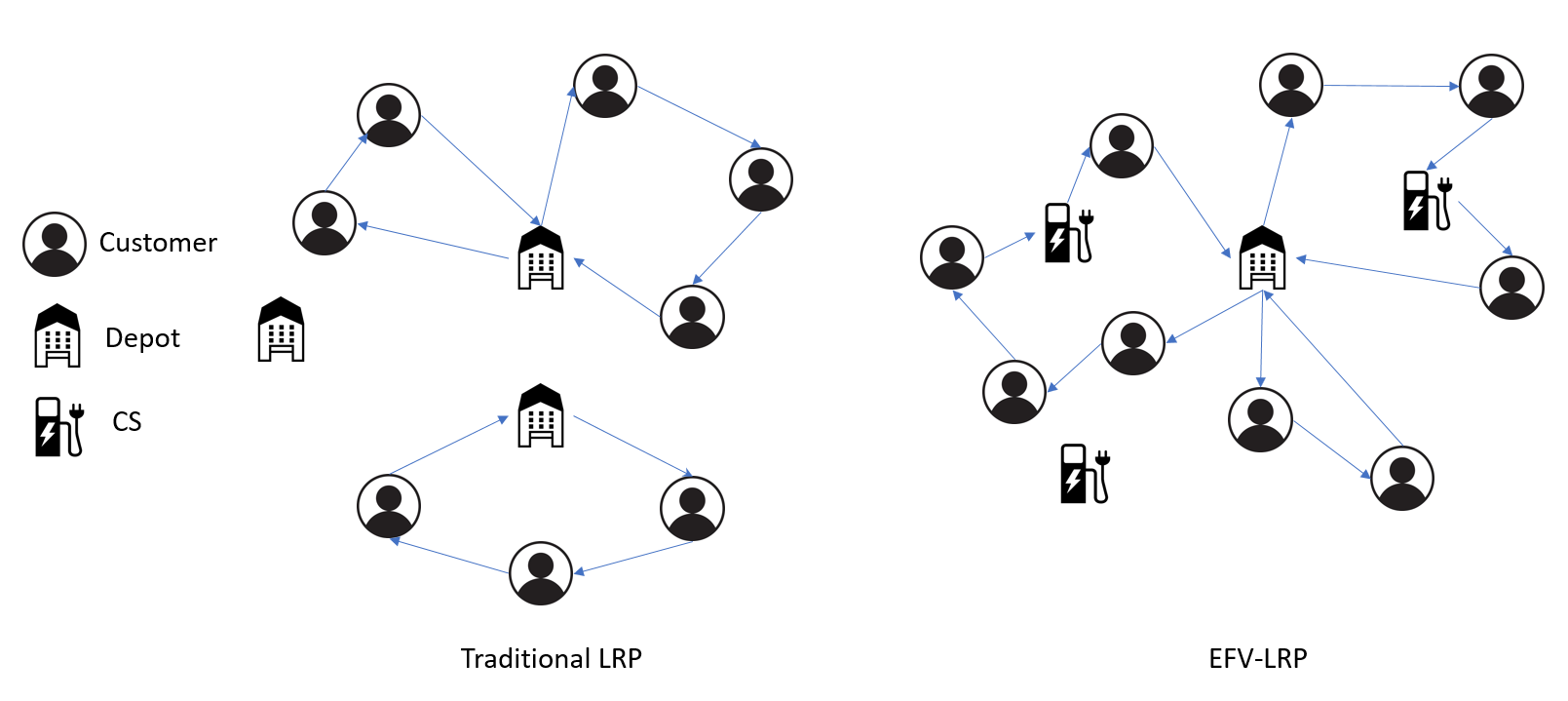}
    \caption{Traditional \ac{LRP} vs \ac{EFV}-\ac{LRP}}
    \label{fig:eLRP}
\end{figure}

In traditional \ac{LRP}, the allocation of depots is decided to serve customers in a particular service area and the routes are designed from different depots to serve customers using a fleet of vehicles. \ac{LRP} can be tailored for \ac{EFV} logistics system by deciding the allocation of \acp{CS} instead of depot allocation, and the route is designed from the depot to serve customers using a fleet of \acp{EFV}, as shown in Figure \ref{fig:eLRP}.  In summary, the goal of the \ac{LRP} is to consider the interaction between planning and operation when optimizing the \ac{CS} locations and the routing of \ac{EFV} simultaneously.

\cite{li2015multiple} proposed an optimization model considering multiple types of \acp{CS} in \ac{LRP} with delivery time window for freight systems. This model was further extended by \cite{schiffer2017electric} to optimize fleet size at the depot. \cite{schiffer2018electric} decided the siting of \ac{CS} and optimal route while minimizing the \ac{TCO} for a given planning period using the data of a German non-food retailer, TEDi. \ac{TCO} includes purchasing costs of \acp{EFV}, installation costs of \ac{CS}, drivers' wage, and operational costs. 

In another study, \cite{schiffer2018strategic} considered the uncertainty in geographical customer distribution in the \ac{LRP} through a robust location-routing approach. \cite{paz2018multi} introduced a multi-depot logistic system for the first time in \ac{LRP} with delivery time windows. The authors assumed that \ac{EFV} departs and returns to the same depot for each shipment. The multi-depot study was further extended to multiple service periods (i.e., different depots have different service times) in \cite{wang2022electric}. 

The service to both linehaul (pickup freight) and backhaul (deliver freight) customers along the route were introduced in \ac{LRP} study by \cite{yang2022integrated}. The heterogeneous fleet of \acp{EFV} considering different freight capacity, battery capacity, and operating cost was introduced in \cite{ccalik2021electric}.  
\cite{wang2021branch} decided the location of depots with either of two energy-restore facilities (i.e., plug-in charging facility or \ac{BSS}) and routing decision simultaneously. \cite{kocc2019electric} studied the \acp{CS} allocation problem, which could be invested and shared by several companies, along with the routing decisions. 

The coupling of \acp{CS} planning and operational scheduling was done for two different mobility patterns, such as contracted mobility pattern and subcontracted mobility pattern \citep{londono2019optimal}. In the contracted mobility pattern, the operation schedule was formulated as \ac{EVRP} (i.e., minimize total traveling cost and consider freight load dynamics), whereas subcontracted mobility was formulated as \ac{SP} problem (i.e., minimize total travel distance only, no consideration of load dynamics).

\section{RQ3: What methodologies have been employed to tackle the planning and operational scheduling for \acp{EFV}?}\label{sec:RQ3}
The existing methodologies of \ac{CS} planning and operational studies can be categorized into four methodologies such as coverage- and queuing-based, mathematical programming, simulation, and survey approaches as shown in Table \ref{tab:methodologies_summary}. Mathematical programming models are most frequently used, particularly in operational scheduling and coupling of location and routing decision problems. However, for \ac{CS} planning coverage-based and queuing-based approaches are popular. As mathematical programming is a popular approach in operational scheduling, the fundamental formulation is also described in section \ref{sec:RQ4}. The extension of such formulation with corresponding logistic patters is described in section \ref{sec:RQ5}.

\begin{table}[H]
    \centering
    \begin{tabular}{|c|c|c|} \hline 
         \multicolumn{2}{|c|}{Methodology}& No. of literature
\\ \hline 
         \multicolumn{2}{|l|}{\textbf{Coverage and queuing model}}& 2
\\ \hline 
         \multicolumn{2}{|l|}{\textbf{Mathematical programming}}& \\ \hline 
         \multicolumn{2}{|l|}{ \textit{MIP}}& 27
\\ \hline 
         \multicolumn{2}{|l|}{\textit{IP}}& 7
\\ \hline 
         \multicolumn{2}{|l|}{\textit{MINLP}}& 4
\\ \hline 
         \multicolumn{2}{|l|}{\textit{LP}}& 2
\\ \hline 
         \multicolumn{2}{|l|}{\textit{NLP}}& 1
\\ \hline 
         \multicolumn{2}{|l|}{\textit{Heuristic}
}& 4
\\ \hline 
         \multicolumn{2}{|l|}{\textit{CP}}& 1\\ \hline 
 \multicolumn{2}{|l|}{\textbf{Simulation}}&8\\ \hline 
 \multicolumn{2}{|l|}{\textbf{Survey}}&1\\ \hline
    \end{tabular}
    \caption{Usage of different methodologies in existing literature}
    \label{tab:methodologies_summary}
\end{table}

\subsection{Coverage-based and queuing-based approach}
The coverage-based approach aims to allocate charging stations to cover a service area. \cite{speth2022public} and \cite{sauter2021charging} decided the siting of \acp{CS} by coverage-based approaches, which defines the candidate charging locations with predefined distance (25 or 50 km in \cite{speth2022public}; 50 or 100 km in \cite{sauter2021charging}). Charging events were estimated heuristically based on a predefined battery range (300 km) and fixed percentage of charging from the public \acp{CS} (assumed 50\% in \cite{speth2022public}; 25\% in \cite{sauter2021charging}) to satisfy the given travel demand. 
\rev{While simple to be implemented, the coverage-based approach relies heavily on the parameter assumptions and cannot be effectively used to determine the optimal sizing of CSs.}

Queuing models can be leveraged to decide the sizing of each \ac{CS}. 
The arrival of \acp{EFV} at a \ac{CS} was typically assumed to follow Poisson distributions. The charging time was assumed to follow an approximate normal distribution in \citep{sauter2021charging} or Markovian distribution in \citep{speth2022public}. The queuing model was solved by approximation formulas (extension of the Pollaczek Khinchine formula) as described in \cite{funke2018techno}. 

\subsection{Mathematical programming model}
\rev{The above-mentioned coverage-based and queue-based approaches provide good guidance for \ac{CS} planning, however, the decisions made are not guaranteed to be system optimal.}
To quantitatively optimize the sitting and sizing decisions of \acp{CS}, mathematical programming approaches can be used. The application of the mathematical programming for allocation of \ac{CS} were proposed in \acp{LRP}.  

In contrast to modeling most of the operational problems using holistic mathematical programming approaches such as \ac{MIP}, \ac{IP}, \ac{MINLP} etc., some studies used heuristic procedures to determine ``good'' routing strategies, which may not be optimal. \cite{montoya2016multi} implemented \ac{MSH} (introduced in \cite{mendoza2013multi}) for \ac{EFV} routing problems. \ac{MSH} has two phases: sampling and assembling. During the sampling phase, routes from the traveling-salesman-problem tours are sampled, whereas the assembling phase uses the set-partitioning formulation. The first phase builds a pool of routes via a set of randomized route-first cluster-second heuristics and the second phase assembles a routing decision by solving a set partitioning formulation over the routes that are stored in the pool. The former controls the feasibility and the cost of each route while the latter controls the feasibility and the cost of the whole solution. \cite{andelmin2019multi} implemented a multi-start local search heuristic that is built based on the idea of a multi-start method proposed in  \cite{marti2013multi}. Among the three phases of the multi-start local-search heuristic, the first two phases iteratively construct new solutions which are further improved by the local search algorithm and finally, all vehicles' routes associated with improved solutions are stored in a route pool. The set of optimal routes from this route pool is decided from a set partitioning algorithm in the third phase, which is further improved by local search algorithms. 

\cite{reyes2019simheuristic} implemented a simheuristic approach (integration of simulation techniques within a heuristic framework to address stochasticity) which consists of two components. First, a meta-heuristic component searches for and stores promising solutions. Second, the simulation component estimates different performance statistics of these solutions considering the stochastic environment. Here, \cite{reyes2019simheuristic} proposed a biased-randomized multi-start framework that integrated \ac{MCS} to reflect the stochasticity of \ac{EFV} travel time. \cite{woody2021charging} studied charging emission in terms of \ac{MEF} considering four charging profiles based on charging \ac{SOC} (full, flexible) and timing of charging at depot (immediate, delayed). \ac{MEF} was estimated with the regression method that was proposed in \cite{donti2019much}. The sensitivity of the \ac{GHG} emission was also studied in \cite{woody2021charging} using a simulation technique that was proposed with details in \citep{deetjen2019reduced}.

\subsection{Simulation-based approach}
\rev{Although mathematical programming is a rigorous quantitative approach for making optimal operational decisions, this approach could be computationally expensive for large-scale problems in the real world and formulating such models may become challenging for very complex systems and introducing the uncertainty. In contrast to the pursuit of optimal decisions, utilizing simulation models can prove to be an effective method for comprehending the intricate dynamics exhibited by a system when the simulation environment can be effectively calibrated.}

For example, after siting \acp{CS} in depots, \cite{borlaug2021heavy} investigated the required upgrades in power distribution systems to serve the charging demand of heavy-duty \acp{EFV} using bootstrap sampling of different scenarios in a simulation-based framework. Considering the installation of \acp{DCFC} at existing gas stations, \cite{bischoff2019impacts} developed a simulation model using MATSim to estimate charging demand. Considering the installation of \acp{CS} at truck stops, \cite{hurtado2021driving} developed a GIS model to study the driving coverage of \acp{EFV}. \cite{jahangir2021road} implemented simulation model ( \acp{BEVPO}) \citep{bevpo} to investigate the range anxiety and potential of \acp{EFV} adoption.  

In context of operational scheduling problem, both commercial and open-source simulation software were used. \cite{napoli2021freight} used commercial transportation planning software TransCAD (which has built-in route optimization solving feature) for \ac{EFV} routing problems. Commercial traffic simulation software VISUM was used to simulate traffic network states (e.g., travel time, speed, etc) along the route in \cite{chen2021mixed} given the route flow calculated from optimization models. \cite{ewert2021using} simulated food retail distribution using open-source agent-based microscopic simulation software MATSim. The jsprit was integrated into MATSim for route optimization purposes. \rev{However, it's worth noting that these software solutions are not specifically designed for \acp{EFV}' operational scheduling, and customization may be needed to  realistic represent the system and achieve optimal operational decisions.}

\subsection{Survey approach}
In addition to quantitative siting and sizing studies, \cite{melander2022benefits} conducted qualitative surveys to compare four business strategies of sharing \acp{CS}, including sharing existing privately owned infrastructure, jointly building new infrastructure at a shared location, jointly building new infrastructure at customers' site, and organizing sharing through a third party, to explore \ac{CS} planning strategies to improve economic and environmental benefits. 

\section{RQ4: How to formulate fundamental mathematical programming model?}\label{sec:RQ4}
\ac{LRP} and operational scheduling problems were found to be formulated as \ac{IP}, \ac{MIP} and \ac{MINLP} models, with single study formulating \ac{MIP} for \ac{CS} planning. Two key components of mathematical programming models such as objective functions and constraints are different for \ac{CS} planning, operational scheduling and coupling of \ac{CS} planning and operations. To enumerate all constraints and objective function components in this review paper will be lengthy and unnecessary. Instead, we will only mention the common and fundamental objective function components and constraints found in the existing literature.
\\
\\
\subsection{Model Objective}

\textbf{\ac{CS} planning}
The deployment and installation costs of CSs with \ac{EVSE} and operational costs are minimized in the objective function of \ac{CS} planning given the unit cost of each CS, EVSE and CS operations \citep{alp2022transitioning,li2023centralized}.

\textbf{Operational scheduling}
Due to limited traveling distance and the range anxiety of \acp{EFV}, a common objective in the existing operational scheduling studies (particularly routing problems) is found to minimize total traveled distance \citep{sadati2021hybrid,andelmin2019multi,sadati2022electric,afroditi2014electric,goeke2019granular,elahi2022modeling,goeke2015routing,yilmaz2022variable,akbay2022variable} or its corresponding energy consumption costs \citep{zhao2021vehicle,PELLETIER2019225,napoli2021freight,scholl2023platooning}. Since en-route charging will significantly delay the freight delivery, minimizing the quantity of en-route charging energy or its corresponding charging cost is also found in the existing studies \citep{pelletier2018charge,alam2022charging,elahi2022modeling,arias2021linearized}. At the depot, the peak load charge could be costly when significant charging activities are conducted at the same time. \cite{al2022smart} proposed a smart charging system at the depot that minimizes the peak demand of the aggregate charging load. There are some studies that converted all cost components into time or \ac{VOT}. For example, traveling time and charging time are minimized in \cite{ahmadi2021vehicle,cortes2019electric} and \ac{VOT} is minimized in  \cite{reyes2019simheuristic,chen2021mixed,van2013towards}. 

In the freight industry, some other costs are also significant due to the unique freight characteristics. For example, due to the higher hourly wage of a truck driver, minimizing truck driver's wage costs including overtime costs have been considered in the existing studies \citep{bac2021optimization,goeke2015routing,PELLETIER2019225,van2013towards,zhao2021vehicle,wang2021integrated}. Due to tight delivery schedule and to ensure customer satisfaction in the competitive freight industry, minimizing delay cost (i.e., the penalty for the delivery delay) is also considered in recent studies \citep{ALAM2023104009,zhao2020distribution,bac2021optimization,zhao2021vehicle}. Compared with traditional \ac{ICEV}, \ac{EFV} is more expensive currently. Therefore, operational studies, especially the studies of fleet sizing problems, minimize \ac{EFV} dispatching cost incurred from purchasing (fixed) or renting (per unit time of usage) vehicles \citep{van2013towards,PELLETIER2019225,zhao2020distribution,hiermann2016electric}. 

\textbf{Coupling \ac{CS} planning and operational scheduling}
Most of the studies on \ac{LRP} minimized costs for installing new \acp{CS}, maintenance of these \acp{CS}, charging, driver's wage, and delay\citep{alp2022transitioning, londono2019optimal,li2015multiple, yang2022integrated,wang2021branch,kocc2019electric}.  
In addition, total traveled distance (corresponds to energy consumption costs) was minimized \citep{ccalik2021electric}. 

\subsection{Constraints}
\textbf{\ac{CS} planning}
A set of constraints were formulated for planning the deployment of \ac{CS} and installation of \ac{EVSE} over the planning horizon. Balance-equation constraints for \acp{EVSE} balanced the purchasing and salvaging of \ac{EVSE}. Another set of constraints was added to ensure the completion of freight delivery. The budget constraints controlled the new investment for \acp{CS} and \acp{EVSE} considering discount factors for different years.

\textbf{Operational scheduling}
\textit{Freight network constraints}:
Freight networks can be modeled as directional graphs. To ensure the realistic representation of freight operation, the following graph constraints are typically imposed (e.g., \citep{yilmaz2022variable,schneider2014electric,ahmadi2021vehicle}). 

\begin{itemize}
    \item In-degree and out-degree constraints define that exactly one edge enters and leaves each node of the freight network associated with each customer, respectively. This constraint ensures that \ac{EFV} departs after visiting a customer or \ac{CS} and flow conservation is guaranteed in routing. 
    \item To run the shipment cycle on a regular basis, \ac{EFV} should return to its origin after completing the deliveries along the route. This property can be ensured by constraining that the number of \acp{EFV} leaving the depot equals the number of \acp{EFV} entering the depot. 
    \item Sub-tour elimination constraint prevents a trip visiting few cities in the given set and return to the starting city.
    \item Each customer is visited exactly once which enforces the connectivity of customer visits so that overall freight transportation is organized and streamlined. 
\end{itemize} 

\textit{Freight load constraints}:
Freight load constraints are crucial for tracking freight load while \ac{EFV} moving along the route and satisfying customer demand. Some typical constraints found in the existing literature (e.g., \cite{yilmaz2022variable,schneider2014electric,ahmadi2021vehicle,arias2021linearized}) are as follows.  
\begin{itemize}
    \item The freight flowing through each edge is less than the freight capacity of the \acp{EFV}.   
    \item The amount of freight to be delivered by an \ac{EFV} after visiting the \ac{CS} remains the same because charging has no impact on freight load \citep{yilmaz2022variable}.
    \item The weight of the \acp{EFV} right after leaving the depot is equal to the sum of the freight weight and \ac{EFV}'s own dead weight. 
\end{itemize}

\textit{Battery \ac{SOC} constraints}:
The constraints for tracking battery \ac{SOC} are fundamental to model the energy dynamics and charging behavior of \acp{EFV}. Dynamics of battery \ac{SOC} can be modeled by the following constraints \citep{van2013towards,pelletier2018charge,yilmaz2022variable} 
\begin{itemize}
    \item The real-time battery energy of each \ac{EFV} shall not exceed its battery capacity. Based on the distance traveled, the battery \ac{SOC} is reduced along the route unless charged. The remaining distance (i.e., the distance that can be traveled by the remaining battery \ac{SOC}) is also monitored. The initial remaining distance is equal to the maximum battery range of the \ac{EFV}. The remaining distance is updated by subtracting the distance traveled between nodes based on MPGe. No trip shall be conducted with a trip distance exceeding the remaining distance of a vehicle. 
    \item  
    Without loss of generality, a constraint is imposed to ensure that the \ac{SOC} needs to be greater than a minimum percentage and below the maximum percentage of battery capacity. 
    For example, \ac{EFV} are not allowed to reach extremely low \ac{SOC} or constantly be fully charged to enhance the longevity of battery performance and/or emulate real-life user behavior.
    However, full charging is assumed at en-route \ac{CS} by some existing studies (e.g., \cite{elahi2022modeling}).
    \item In charging scheduling problem, each \ac{EFV} will occupy at most one charger per period \citep{pelletier2018charge}. Such constraint is needed to determine the charging availability and charger allocation.
    \item  To address the variable charging speed based on different battery \ac{SOC}, a charger-specific piece-wise linear function needs to be implemented to formulate a \ac{LP} model. 
    \item Majority of the studies assume that \acp{EFV} leave the depot fully charged. A constraint is required to ensure this initial condition.
    \item Considering smart charging with \ac{V2G} capability where bi-directional power flow is considered, in the real-time operational scheduling mode, the lower bound of charged power (i.e., the upper bound of discharged power) at a particular timestamp for an \ac{EFV} is determined by the availability of the \ac{EFV} at the CS and the maximum discharge power at that particular timestamp. On the other hand, the upper bound of charge power is determined by the availability of the \ac{EFV} at the CS and the maximum charge power \citep{al2022smart}.
\end{itemize}

\textit{Fleet sizing constraints}:
For the fleet sizing problem, the number of utilized \acp{EFV} per period can not exceed the available number of \acp{EFV} \citep{elahi2022modeling,yeomans2021multicriteria,ioannou2020freight}. In addition, the total number of \acp{EFV} allocated for an OD pair should be greater or equal to the demand of the OD pair \citep{chen2021mixed}.

\textit{Platoon constraints}:
The total of \acp{EFV} that join a platoon may not exceed the maximum platoon capacity and at least two \acp{EFV} are required to form a platoon \citep{ALAM2023104009}. When two or more \acp{EFV} form a platoon at any stopping point, their respective departure times should be the same. 
When considering the vehicle sequence in the platoon (e.g., a leading vehicle and following vehicles), constraints are also needed to indicate the position of each vehicle \citep{scholl2023platooning} 

\textbf{Coupling \ac{CS} planning and operational scheduling}
The majority of the constraints are found similar to mathematical programming in \ac{CS} planning and operational scheduling. A new graph constraints, in a multi-depot-based freight network, is that each route is constrained to include only one depot \cite{wang2021branch}. Given the consideration of multiple service periods of the depot, the sequence of routes that are performed by each \ac{EFV} needs to be maintained in \ac{LRP} \citep{wang2022electric}. These constraints are not unique constraints of \ac{LRP}, therefore, these can also be valid for operational scheduling problems in the context of multi-depot-based freight network.

\section{RQ5: How do different logistics patterns give rise to new challenges}\label{sec:RQ5}
We introduce different freight delivery patterns found in the existing literature in the following subsections to facilitate the discussion of operational problems. Different types of logistic patterns cause new challenges to adjust the methodologies accordingly for proper planning and operational decisions. In existing \ac{EFV} literature, these adjustments due to unique logistic pattern in fundamental model are also mentioned after the introduction of each pattern. 

Freight delivery patterns considered in the existing \ac{EFV} literature can be classified into five categories, including heterogeneous customers with and without due time, flexible delivery location, non-simultaneous pickup and delivery (i.e., dedicated-service logistics), simultaneous pickup and delivery and two-echelon delivery. 

\subsection{Heterogeneous customers with and without due time}
\begin{figure}[H]
    \centering
    \begin{subfigure}[tbh!]{.5\linewidth}
	\centering
	\includegraphics[width=1\linewidth]{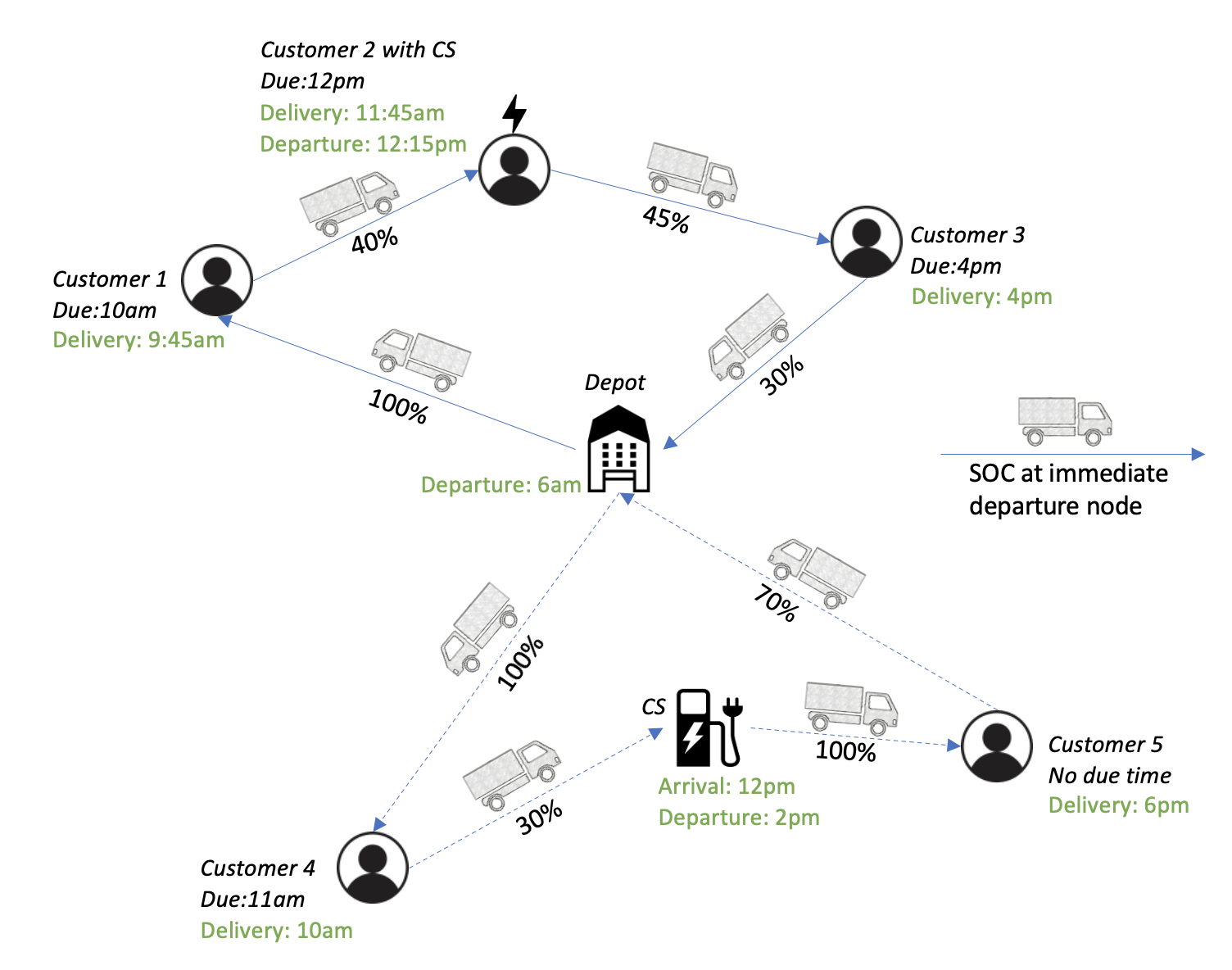}
	\caption{PREV-DT}
	\label{fig:PREV-DT}
    \end{subfigure}\hfil
    \medskip
    \begin{subfigure}[tbh!]{0.4\linewidth}
	\centering
	\includegraphics[width=1\linewidth]{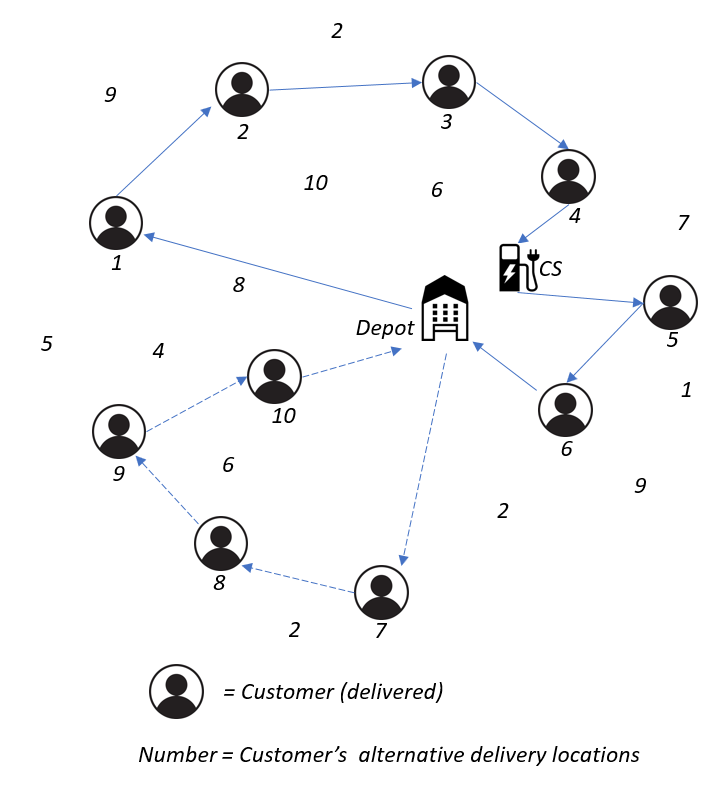} 
	\caption{EVRP-FD}
	\label{fig:EVRP-FD}
    \end{subfigure}
    \medskip
    \begin{subfigure}[tbh!]{.3\linewidth}
	\centering
	\includegraphics[width=1\linewidth]{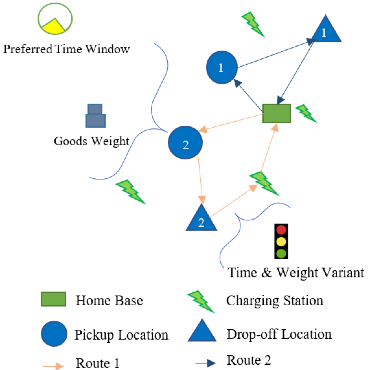}
	\caption{EVRP-PD-TW-PR}
	\label{fig:EVRP-PD-TW-PR}
    \end{subfigure}\hfil
    \medskip
    \begin{subfigure}[tbh!]{0.6\linewidth}
	\centering
	\includegraphics[width=1\linewidth]{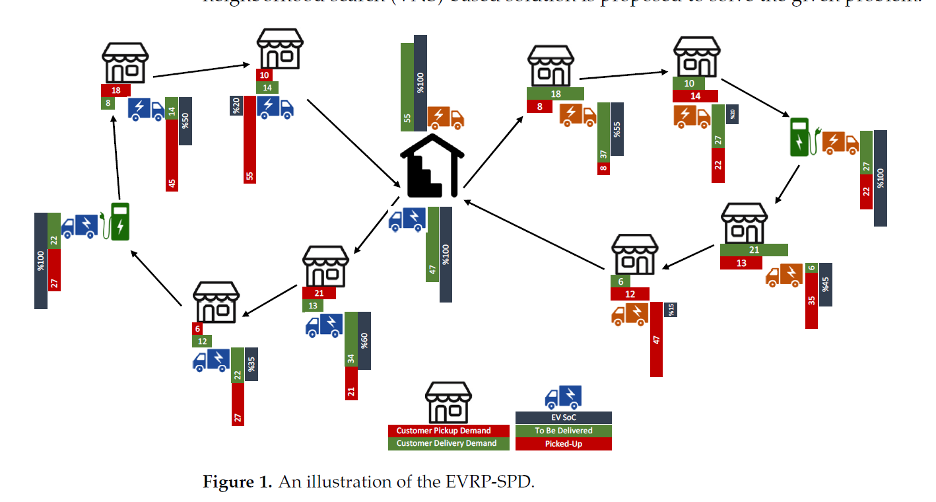} 
	\caption{EVRP-SPD}
	\label{fig:EVRP-SPD}
    \end{subfigure}
    \caption{Popular logistic patterns frequently implemented in EFV studies}
    \label{fig:logistic_patterns}
\end{figure}

Figure \ref{fig:PREV-DT} illustrates \ac{PREV-DT} where customers’ orders are due for delivery at a scheduled time period. Some customers must receive their orders prior to the delivery due time, while others can wait until the end of the period. In Figure \ref{fig:PREV-DT}, we illustrate two \acp{EFV} serving two different routes, which are plotted with solid and dash line types, respectively. Both \acp{EFV} depart from the depot initially and need to finish their delivery tasks along the routes and return to the depot. Each route has different charging infrastructure accessibility and different customers have different delivery due times. The \ac{EFV} serving the solid-lined route can access to onsite charging infrastructure at customer ``2'' and recharge some energy amount so that it can deliver customer ``3" in time given that customer ``3" has delivery due time. On the other hand, the second \ac{EFV} chooses to fully recharge the battery due to no urgency given that customer ``5" does not have a delivery due time.

In order to avoid any penalty or delay cost, customers with due time need to be served within the delivery time window \citep{akbay2022variable,zhao2021vehicle,al2022smart,hiermann2016electric,napoli2021freight,tacs2021electric}. These constraints are crucial for routing problems and can be modeled as hard constraints to guarantee compliance with the time windows.

\subsection{Customers with flexible delivery locations}
Figure \ref{fig:EVRP-FD} illustrates the flexible delivery locations with 10 customers and a total of 23 alternative delivery locations in the system. The scattered numbers represent the index of customers and the same numbers represent the alternative delivery locations of that customer. For instance, customer ``2" has four scattered alternative delivery locations. EFV ``1" (traveling on the solid-line route) departs from the depot, visits customers ``1",``2",``3",``4", and stops at \ac{CS}. After recharging, it resumes by visiting customers ``5" and ``6", and then returns to the depot at the end of its tour. Similarly, EFV ``2" (traveling on the dashed-line route) departs from the depot, serves customers ``7" through ``10", completes its tour, and returns to the depot without recharging.

\subsection{Non-simultaneous (Dedicated-service logistics) pickup and delivery}
Two types of freight pickup and delivery patterns such as non-simultaneous (i.e., dedicated-service logistics) and simultaneous pickup and delivery were found in the existing studies of operational scheduling problems. Figure \ref{fig:EVRP-PD-TW-PR} illustrates non-simultaneous pickup and delivery (i.e., dedicated-service logistics) with en-route flexible charging (i.e., flexible charging amount) given pickup and delivery time windows. \footnote{The ``time windows'' refer to the earliest and latest time range of events, such as departure or delivery.} Assume a set of customers that have scheduled for pickup and delivery with specific time windows as shown in circles and diamonds, respectively. The numbers of the locations denote the customer indexes. At the depot, a \ac{CS} is deployed for overnight charging. The \acp{EFV} should start from and return to the depot within the scheduled hours. There could be other en-route \acp{CS} in the system. To extend the range of \acp{EFV}, flexible charging en-route is allowed. In this figure, two \acp{EFV} depart from the home base (e.g., depot) and serve customers ``1'' and ``2'', respectively. The en-route charging is not needed for serving customer ``1", whereas en-route charging is needed for serving customer ``2". 

In dedicated-service logistics, an additional constraint is needed so that an identical \ac{EFV} will pick up and deliver the freight for the same customer, and the pick-up should be finished before the delivery for the same customer \citep{zhao2021vehicle}.

\subsection{Simultaneous pickup and delivery}
\ac{SPD} is a generalization of non-simultaneous pickup and delivery by allowing flexibility to both pickup and delivery of freights at each customer location. Figure \ref{fig:EVRP-SPD} illustrates the \ac{SPD} freight logistics pattern. Every \ac{EFV} has been labeled by three parameters - ``the amount of freights to be delivered/the amount of freights being picked up/the current \ac{SOC} of the \ac{EFV}''. The amount of freight in the \ac{EFV} changes depending on the pickup and delivery requests by the customers. For example, an \ac{EFV} (starts with 100\% \ac{SOC}) needs to deliver 55 units of freights and pickup packages along the solid-lined route. To finish these delivery and pickup tasks, it recharges the battery at en-route \acp{CS} to its full capacity. When this \ac{EFV} returns to the starting point of the route (e.g., depot), it delivered all of the freights and picked up 47 units from the customers to whom the freights are delivered. After arriving at the home base, this \ac{EFV} will then deliver 47 units of packages to another set of customers along the dash-lined route.

In contrast to freight vehicles for delivery only, vehicles in \ac{SPD} needs additional constraint so that freight flowing through each edge does not exceed the freight capacity of the \acp{EFV} after adjustment of simultaneous pick up and delivery. For \ac{SPD}, freight weight leaving the depot is zero because no freight will be picked up by \acp{EFV} from the depot to customers.

\subsection{Two-echelon delivery}
In two-echelon delivery, first echelon refers to the freight transportation from the central warehouse to satellites (i.e., the last stop of middle-mile trip) by \ac{ICEV} trucks and the second echelon refers to the transportation from satellites to customers using \acp{EFV} \cite{akbay2022variable}. The variants of routing problem from this logistic pattern is called \ac{2E-EVRP-TW}. The charging event can be initiated at satellites (first-echelon) and an alternative transportation mode (e.g., walking, drones, bikes, etc.) for freight delivery to close-distanced customers during charging (second echelon) \cite{cortes2019electric}.

The equality constraint of freight to be delivered by an \ac{EFV} before and after visiting the \ac{CS} is relaxed in a two-echelon delivery where an alternative transportation mode is assigned to deliver freight to nearby customers when charging \acp{EFV}.

\section{RQ6: Which solution approaches have been utilized in addressing mathematical programming? How do they differ?}\label{sec:RQ6}

In small-scale case studies, exact methods were used by calling commercial global optimal solvers, such as Gurobi and CPLEX \citep{arias2021linearized,pelletier2018charge,ALAM2023104009,bac2021optimization,cortes2019electric,akbay2022variable}. However, for the large-scale problems, traditional exact methods could be time-consuming to obtain a global optimal solution due to NP-hardness. Meta-heuristic and hybrid of multiple meta-heuristic algorithms can be used to obtain solutions of reasonable quality for practical purposes. \rev{Choosing meta-heuristic algorithms and determining how to hybridize them is contingent upon the specific nature of the problems, available computational resources, and desired levels of solution quality. However, limited studies has compared the computational performance of different computational algorithms based on standardized test cases, which is a valuable future research direction.}

\begin{table}
    \centering
    \begin{tabular}{|c|c|c|} \hline 
         \multicolumn{2}{|c|}{Mathematical programming solving approach}& No. of literature
\\ \hline 
         Solver&  CPLEX& 4
\\ \hline 
         &  Gurobi& 3
\\ \hline 
         &  chuffed& 1
\\ \hline 
         &  jsprit& 1
\\ \hline 
         &  shortec& 1
\\ \hline 
         &  TransCAD built-in solver& 1
\\ \hline 
         &  DICOPT& 1
\\ \hline 
         Decomposition&  Branch-and-price& 3
\\ \hline 
         &  Benders decomposition& 1
\\ \hline 
 & LR-ADMM decomposition &1
\\ \hline 
 & Column generation &1
\\ \hline 
 Heuristics& Heuristics&6
\\ \hline 
 Meta-heuristics& Adaptive LNS&6
\\ \hline 
 & SAA&2
\\ \hline 
 & Adaptive ACA&1
\\ \hline 
 & BA&1
\\ \hline 
 & GA&1
\\ \hline 
 & LNS&1
\\ \hline 
 & ACA&1
\\ \hline 
 & CW&1
\\ \hline 
 & DP&1
\\ \hline 
 & VNS&1
\\ \hline 
 Hybrid meta-heuristics& VNS-TS&2
\\ \hline 
 & Adaptive VNS-TS&1
\\ \hline 
 & ALNS-DP&1
\\ \hline 
 & CW-Adaptive VNS&1
\\ \hline 
 & CW-VNS&1
\\ \hline 
 & CW-IG-ALNS&1
\\ \hline 
 & GMCA-INSGA-II&1
\\ \hline 
 & Iterated Local Search-VND-Set partitioning&1\\ \hline
    \end{tabular}
    \caption{Solving approaches implemented for mathematical programming models in prior studies}
    \label{tab:math_programming_solve_approach}
\end{table}

Among nature-based meta-heuristic algorithms, \ac{GA} was implemented in \citep{wang2021integrated} in which the study problem had two parts: first, use \ac{CBR} to forecast energy consumption and travel time of each route by implementing neural network and fuzzy inference mechanism; second, optimize vehicle routing by implementing \ac{GA} using the forecast energy consumption and travel time. \rev{In contrast to \ac{GA}, which can only be solved centrally}, \ac{ACA} is well suited for distributed computing, 
which makes itself compatible to combine with other meta-heuristic algorithms. \ac{ACA} and adaptive \ac{ACA} was implemented in \cite{zhao2021vehicle} and \cite{zhao2020distribution}, respectively, for the routing problem. \ac{BA} is another nature-based meta-heuristic, which is adopted in \cite{yeomans2021multicriteria} to produce multiple alternative routing solutions. 

\rev{In addition to nature-based meta-heuristic algorithms, there exists another category of meta-heuristics based on local search. While nature-based algorithms navigate the search space using a predefined population to discover high-quality solutions, local-search-based algorithms begin with an initial solution and focus on seeking improved solutions within a defined neighborhood. However, some local search heuristics may not achieve a high-quality solution, as they prioritize time-efficient solutions that balance optimality with achieving satisfactory outcomes.}

Local-search-based meta-heuristic algorithms typically combine perturbation and greedy approaches to escape local minimum. For example, \ac{SAA} tries to overcome the dependency of initial value and being stuck on local optimal solutions by utilizing a probability function (i.e., metropolis criterion) which allows ``locally'' worse solutions to be selected during the optimization process \citep{elahi2022modeling}. Besides, \ac{SAA} has desired properties, such as robustness, flexibility, global search, and ease of implementation.\ac{SAA} was implemented for routing problems \citep{fan2022half,elahi2022modeling}. \ac{VNS} algorithm is a combination of the descent method that moves to a local minimum, and the shaking method that escapes from being trapped in a local minimum \citep{bac2021optimization}. \cite{yilmaz2022variable,akbay2022variable} implemented \ac{VNS} leveraging the initial solution using \ac{CW}  saving algorithm. \ac{VND} is deterministic version of \ac{VNS} that was implemented by \cite{bac2021optimization} for the local search procedure leveraged solution from \ac{VNS}. An adaptive general \ac{VNS} was also implemented for milk collection problem in \cite{erdem2023optimisation}.  
Compared with \ac{VNS}, another local-search meta-heuristic algorithm \ac{LNS} \citep{PELLETIER2019225} has a high possibility to escape local optima and explore new regions of the search space, because if there is no yield of improvements from local search, \ac{LNS} applies a large-scale perturbation through destroy-repair approach (the destroy component destructs some parts of the current solution while the repair method rebuilds the destroyed solution) to the current solution. The extension of \ac{LNS}, adaptive \ac{LNS} was found popular to be implemented in the existing literature \citep{hiermann2016electric,goeke2015routing}. Unlike \ac{LNS}, adaptive \ac{LNS} is not restricted to one destroy and repair heuristic rather, it chooses solutions in every iteration from a pool of heuristics based on past success \citep{lutz2015adaptive}.  

\rev{In recent years, the hybridization of meta-heuristic algorithms has become popular due to its ability to combine the strengths and complementary features of different meta-heuristic algorithms to create more powerful optimization algorithm.} \cite{sadati2021hybrid,sadati2022electric} proposed a hybrid algorithm of \ac{VNS} and \ac{TS} in which \ac{TS} was implemented within \ac{VNS} for solving problem-specific neighborhood structures as part of the local search mechanism \citep{sadati2021hybrid,sadati2022electric}. It is found that the hybrid of \ac{VNS} and \ac{TS} outperforms adaptive \ac{VNS} and \ac{MSH} in terms of computation time with an acceptable solution. When the problem consists of several routing problems, iterated local search meta-heuristic framework is a good fit. This framework generates a set of local optimal routes in \cite{cortes2019electric} by solving \ac{MIP} model using \ac{VND} algorithm (alternating local search and perturbation of customer sets). Once the iterated local search is done, the set partitioning (\ac{IP}) optimization model is solved to find optimal routes. 

\rev{In addition to hybridizing meta-heuristic algorithms, there are instances where hybrids of exact algorithms such as branch
and bound, and decomposition methods (e.g., column generation) are also implemented. }  \cite{klein2022branch,klein2022integrated,hiermann2016electric} implemented branch and price algorithm to solve large-scale case studies of charging scheduling for \acp{EFV}. 
Compared with meta-heuristic algorithms, a hybrid of exact algorithms guarantee global convergence theoretically, but the convergence rate can be slow for some larger-scale cases. Combining meta-heuristic and exact algorithms is also possible to identify good initial solutions first using meta-heuristic algorithms and then use exact algorithms to close the optimality gap.

Besides, other tools/solvers that were used for routing problems are ``Chuffed'' \citep{ahmadi2021vehicle}, ``jsprit'' \citep{ewert2021using}, and ``Shortec'' \citep{van2013towards}. \textit{Chuffed} is a \ac{CP} solver based on lazy clause generation. A lazy clause generation is a hybrid approach to constraint solving that combines features of finite domain propagation and boolean satisfiability \citep{chu2018chuffed}. Note that compared to \ac{MIP}, \ac{CP} provides a greater degree of flexibility when non-linear
constraints are handled. The Java-based open-source solver \textit{jsprit} is dedicated to solve traveling salesman and vehicle routing problems. Besides open-source solvers, commercial software such as \textit{Shortec} was also implemented for \acp{EFV} routing problems that use sequential insertion heuristics to solve the problems \citep{GraphHopper}. 

In small-scale case studies for \ac{LRP}, exact methods were used for a globally optimal solution using commercial solvers such as Gurobi \citep{ schiffer2017electric}, CPLEX \citep{li2015multiple,paz2018multi,londono2019optimal} and DICOPT \citep{londono2019optimal}. For large-scale problems, directly leveraging commercial solvers could be time-consuming to obtain a globally optimal solution. Therefore, decomposition algorithms (e.g., bender's decomposition \citep{ccalik2021electric}) and a hybrid of exact algorithms (e.g., branch and price \citep{wang2021branch}) were implemented to solve large-scale case studies. A hybrid of \ac{LR} and \ac{ADMM} decomposition framework was proposed by \cite{yang2022integrated} to decouple the \ac{LRP} into a recharging station location problem and an electric vehicle routing problem. The solution quality was evaluated by the optimality gap for each iteration. 

Due to complexity and scaling issues, similar to operational scheduling studies, many \ac{LRP} studies also proposed meta-heuristic, and hybrid meta-heuristic algorithms for approximate solutions. For example, \cite{kocc2019electric} implemented \ac{LNS} algorithm and introduced several new operators, as well as initialization, intensification, and diversification procedures. The hybrid of adaptive \ac{LNS} and \ac{DP} was found effective compared with exact solver solutions for small-scale instances \citep{schiffer2016ecvs,schiffer2018electric,schiffer2018strategic}. Similarly, a hybrid of adaptive \ac{VNS} and \ac{TS} was compared with exact solutions of small-scale instances in \citep{li2015multiple} where \ac{TS} was implemented for intensification mechanisms. A \ac{DP}-based heuristic, which combines restricted \ac{DP} and Prim’s algorithms was implemented to solve larger instance of dynamic routing optimization problem with uncertainties \citep{unal2022dynamic}. \cite{wang2022electric} introduced a hybrid of \ac{GMCA} and \ac{INSGA-II} for the proposed \ac{LRP}, where \ac{GMCA} assigns customers with depots in various service periods to reduce computational complexity and \ac{INSGA-II} obtains the Pareto optimal solution. This hybrid approach was also compared with multi-objective \ac{GA}, \ac{PSA}, and \ac{ACA}. In another study, \cite{guo2022simultaneous} solved the \ac{LRP} considering non-linear charging profile and battery degradation using a hybrid of three algorithms such as \ac{CW}, \ac{IG} and adaptive \ac{LNS}.

\section{Conclusions and RQ7: What are the potential research areas and challenges that have not been adequately explored}
\label{sec:RQ7}
Transportation electrification in the freight sector is a possible trend in the near future. This paper reviews the existing research on the challenges and opportunities of adopting electric vehicles in freight transportation. In particular, we focus on three major themes, including \ac{CS} planning, \acp{EFV} operational scheduling, and the coupling of \ac{CS} planning and operation, with a detailed classification of the proposed methodologies and solution approaches. Based on our review, we identified further research directions that can further facilitate freight electrification, as summarized below.

\begin{enumerate}

    \item The existing research typically focuses on centralized decision-makers to optimize the planning and operational decisions. However, these decisions can involve multiple stakeholders, such as charging station providers, power utilities, and freight companies, who make decisions in a decentralized manner to optimize their own objectives but whose decisions will interact with each other. The application of a game theoretical modeling framework can be an effective strategy to better understand their decentralized interaction and derive effective policy incentives to guide the system evolvement for societal benefits. 

    \item The existing literature mainly focuses on the planning and operation of \acp{EFV} considering transportation systems. However, freight electrification may have significant impacts on both power transmission and distribution systems due to higher charging power and larger battery capacity. In addition to planning charging stations, a necessary power system upgrade may be needed to support the additional charging power. Furthermore, considering \ac{V2G}, \acp{EFV} fleet can be valuable distributed energy resources to support power system operation compared with passenger vehicles because of larger battery capacity and more predictable travel and charging patterns. 
    
    \item Most of the studies focus on planning and operation of \acp{EFV} for normal scenarios considering economic efficiency. However, freight electrification could face significant challenges during disruptions due to the urgency of good delivery and potential charging inaccessibility due to power or transportation disruptions. More studies will be needed to consider the challenges and opportunities of \acp{EFV} during emergencies to enhance system reliability and resilience.

    \item From a modeling perspective, there are some critical features that can be further considered to shed light on the efficiency and feasibility of adopting \acp{EFV}, including the uncertainties of charger/driver availability, the influence of charging/electricity prices, human-drive v.s. automated, special types of logistics (e.g., logistics for humanitarian, human organs, refrigerated goods, etc), energy consumption factors (e.g., road pavement condition \citep{alam2020systematic}, seasonal and geospatial impact \citep{hao2020seasonal}), and land use restriction. 

    \item Although existing studies have proposed different computational algorithms, there is still a lack of understanding on the comparison of computational performance and solution qualities among different solution approaches. A comparison using standardized models and benchmark test systems for different type of problems will be beneficial.

    \item Solutions of planning/operational decision models can be validated through field experiments (e.g., pilot studies). Validated results from pilot studies will further encourage more stakeholders in freight systems to consider \acp{EFV} in their decision-making process and promote \acp{EFV} adoption. 
    
\end{enumerate}

\section*{ABBREVIATIONS}
\begin{acronym}[ICANN]
    \acro   {GHG}   [GHG]   {Greenhouse Gas}
    \acro   {2E-EVRP-TW}  [2E-EVRP-TW]    {Two Echelon  Electric Vehicle Routing Problem with Time Windows}
    \acro{ACA}[ACA]{Ant Colony Algorithm}
    \acro{ADMM}[ADMM]{Alternating Direction Method of Multipliers}
    \acro{BA}[BA]{Bat Algorithm}
    \acro{BEVPO}[BEVPO]{Battery Electric Vehicle POtential model}
    \acro{BSS}[BSS]{Battery Swapping Station}
    \acrodefplural{BSS}[BSSs]{Battery Swapping Stations}
    \acro{CBR}{Case-Based Reasoning}
    \acro{CS}[CS]{Charging Station}
    \acrodefplural{CS}[CSs]{Charging Stations}
    \acro{CP}[CP]{Constraint Programming}
    \acro{DP}[DP]{Dynamic Programming}
    \acro{EV}[EV]{Electric Vehicle}
    \acrodefplural{EV}[EVs]{Electric Vehicles}
    \acro{CW}[CW]{Clarke and Wright}
    \acro{DCFC}[DCFC]{Direct Current Fast Charger}
    \acro{EFV}[EFV]{Electric Freight Vehicle}
    \acrodefplural{EFV}[EFVs]{Electric Freight Vehicles}
    \acro{eLCVRP-CS-PDS}[eLCVRP-CS-PDS]{Electric Light Commercial Vehicles Routing Problem with Charging Station Location and impact on Power Distribution System}
    \acro{EVRP}[EVRP]{Electric Vehicle Routing Problem}
    \acro{EVRP-ECU}[EVRP-ECU]{Electric Vehicle Routing Problem with Energy Consumption Uncertainty}
    \acro{EVRP-FD}[EVRP-FD]{Electric Vehicle Routing Problem with Flexible Deliveries}
    \acro{EVRP-PD}[EVRP-PD]{Electric Vehicle Routing Problem with Pickup and Delivery}
    \acro{EVRP-PD-TW-PR}[EVRP-PD-TW-PR]{Electric Vehicle Routing Problem with Pickup and Delivery, Time Windows, and Partial Recharge}
    \acro{EVRP-ST}[EVRP-ST]{Electric Vehicle Routing Problem with Stochastic Travel times}
    \acro{EVRPTW}[EVRPTW]{Electric Vehicle Routing Problem with Time Windows}
    \acro{EVRPFTW}[EVRPFTW]{Electric Vehicle Routing Problem with Flexible Time Windows}
    \acro{EVRPTWsc}[EVRPTWsc]{Electric Vehicle Routing Problem with Time Windows and satellite customers}
    \acro{EVRP-TTC}[EVRP-TTC]{Electric Vehicle Routing Problem with Time-varying Traffic Conditions}
    \acro{EVRP-SPD}[EVRP-SPD]{Electric Vehicle Routing Problem with Simultaneous Pickup and Delivery}
    \acro{EVSE}[EVSE]{Electric Vehicle Supply Equipment}
    \acrodefplural{EVSE}[EVSE]{Electric Vehicle Supply Equipments}
    \acro{GA}[GA]{Genetic Algorithm}
    \acro{GMCA}[GMCA]{Gaussian Mixture Clustering Algorithm}
    \acro{GTS}[GTS]{Granular Tabu Search}
    \acro{GVWR}[GVWR]{Gross Vehicle Weight Ratings}
    \acro{HOTDMD-EVRP-BRS}[HOTDMD-EVRP-BRS]{Half-Open Time-Dependent Multi-Depot Electric Vehicle Routing Problem Considering Battery Recharging and Swapping}
    \acro{IBDR}[IBDR]{Incentive-Based Demand Response}
    \acro{ICEV}[ICEV]{Internal Combustion Engine Vehicle}
    \acrodefplural{ICEV}[ICEVs]{Internal Combustion Engine Vehicles}
    \acro{IG}[IG]{Iterative Greedy algorithm}
    \acro{IP}[IP]{Integer Linear Programming}
    \acro{INSGA-II}[INSGA-II]{Improved Nondominated Sorting Genetic Algorithm-II}
    \acro{LRP}[LRP]{Location Routing Problem}
    \acro{LNS}[LNS]{Large Neighborhood Search}	
    \acro{LP}[LP]{Linear Programming}
    \acro{LR}[LR]{Lagrangian relaxation}
    \acro{MSH}[MSH]{Multi-space Sampling Heuristic}
    \acro{MVRPTW}[MVRPTW]{Mixed vehicle routing problem with Time Windows}	
    \acro{MCS}[MCS]{Monte Carlo Simulation}
    \acro{MEF}[MEF]{Marginal Emission Factor}
    \acro{MDGVRP}[MDGVRP]{Multi-Depot Green Vehicle Routing Problem}
    \acro{MIP}[MIP]{Mixed Integer Linear Programming}
    \acro{MINLP}[MINLP]{Mixed Integer Non-Linear Programming}
    \acro{OSPT}[OSPT]{Operational Schedule during Parking Time}
    \acro{PREV-DT}[PREV-DT]{Periodic Routing of Electric Vehicles with Due Time}
    \acro{PSA}[PSA]{Paricle Swarm Algorithm}
    \acro{QP}[QP]{Quadratic Programming}
    \acro{SAA}[SAA]{Simulated Annealing Algorithm}
    \acro{SOC}[SOC]{State Of Charge}
    \acro{SP}[SP]{Shortest Path}
    \acro{SPD}[SPD]{Simultaneous Pickup and Delivery}
    \acro{TS}[TS]{Tabu Search}
    \acro{TCO}[TCO]{Total Cost of Ownership}
    \acro{VOT}[VOT]{Value Of Time}
    \acro{VMT}[VMT]{Vehicle Miles Travelled}
    \acro{VND}[VND]{Variable Neighborhood Descent} 	 
    \acro{VNS}[VNS]{Variable Neighborhood Search}
    \acro{VRP}[VRP]{Vehicle Routing Problem}
    \acro{V2G}[V2G]{Vehicle To Grid}
    \acro{Univ.}[Univ.]{University}
    \acro{TR Part E}[TR Part E]{Transportation Research Part E}
    \acro{Sci.}[Sci.]{Science}
    \acro{COR}[COR]{Computers and Operations Research}
    \acro{WSEAS}[WSEAS]{World Scientific and Engineering Academy and Society}
    \acro{TCS}[TCS]{Transactions on Circuit and Systems}
    \acro{SORT}[SORT]{Statistics and Operations Research Transactions}
    \acro{Idescat}[Idescat]{Statistical Institute of Catalonia}
    \acro{TR Part B}[TR Part B]{Transportation Research Part B}
    \acro{EST}[EST]{Environmental Science and Technology}
    \acro{TR Part C}[TR Part C]{Transportation Research Part C}
    \acro{MPE}[MPE]{Mathematical Problems in Engineering}
    \acro{EJOR}[EJOR]{European Journal of Operational Research}
    \acro{IJPR}[IJPR]{International Journal of Production Research}
    \acro{TRP}[TRP]{Transportation Research Procedia}
    \acro{TRENG}[TRENG]{TRansportation ENGineering}
    \acro{NCST}[NCST]{National Center for Sustainable Transportation}
    \acro{IJIEC}[IJIEC]{International Journal of Industrial Engineering Computation}
    \acro{T-ITS}[T-ITS]{Transactions on Intelligent Transportation Systems}
    \acro{ESPR}[ESPR]{Environmental Science and Pollution Research}
    \acro{SCS}[SCS]{Sustainable Cities and Society}
    \acro{BSE}[BSE]{Business Strategy and the Environment}
    \acro{ERIS}[ERIS]{Environmental Research: Infrastructure and Sustainability}
    \acro{ITOR}[ITOR]{International Transactions in Operational Research}
    \acro{Transp. Lett.}[Transp. Lett.]{Transportation Letters}
    \acro{JESTR}[JESTR]{Journal of Engineering Science \& Technology Review}
    \acro{EUT}[EUT]{Edizioni Università di Trieste}
    \acro{SCLCI}[SD-LCI]{Schloss Dagstuhl – Leibniz Center for Informatics}
    \acro{IOP}[IOP]{Institute Of Physics}
    \acro{Fraunhofer ISI}[Fraunhofer ISI]{Fraunhofer-Institut für System- und Innovationsforschung}
    \acro{EMaTTech}[EMaTTech]{Eastern Macedonia and Thrace institute of Technology}
\end{acronym}
\bibliography{ref.bib}
\pagebreak 
\appendix
%
\pagebreak 

\end{document}